# High-power all-fiber ultra-low noise laser

Jian Zhao,[1, 6] Germain Guiraud,[1, 2] Christophe Pierre,[3] Florian Floissat,[1] Alexis Casanova,[1,5] Ali Hreibi,[4] Walid Chaibi,[4] Nicholas Traynor,[2] Johan Boullet,[3] and Giorgio Santarelli[1,*]

**Abstract**

High-power ultra-low noise single-mode single-frequency lasers are in great demand for interferometric metrology. Robust, compact all-fiber lasers represent one of the most promising technologies to replace the current laser sources in use based on injection-locked ring resonators or multi-stage solid-state amplifiers. Here, a linearly-polarized high-power ultra-low noise all-fiber laser is demonstrated at a power level of 100 W. Special care has been taken in the study of relative intensity noise (RIN) and its reduction. Using an optimized servo actuator to directly control the driving current of the pump laser diode (LD), we obtain a large feedback bandwidth of up to 1.3 MHz. The RIN reaches -160 dBc/Hz between 3 kHz and 20 kHz.

## 1 Introduction

Low-noise high-power single-frequency lasers provide a versatile instrument for fundamental research in interferometric gravitational wave (GW) detectors [1–3], ultra-cold atom/molecular cooling and trapping [4, 5], and precision time/frequency metrology [6]. State-of-the-art low-noise laser sources are usually realized using bulk crystals as gain medium and free-space optics [7–10]. The inherent drawbacks of this approach include stringent optical alignment requirements, a lack of robustness, a large footprint, and high costs. Optical fibers are the ideal gain medium to obtain laser systems with high output power, robust, reliable operation and compact integration. In the single-frequency regime, the master oscillator power amplifier (MOPA) is often preferred. In this configuration, a low-power and low-noise (<1 W) laser is amplified by gain stages to the desired output power. However, a power-scaling limitation for fiber amplifiers in single-frequency regime arises due to the onset of stimulated Brillouin scattering (SBS). Several methods can be deployed to increase the power level at which SBS appears, such as applying fiber strain [14, 16] and thermal gradients [11]. Unfortunately, these approaches are typically incompatible with reproducible and robust operation. Lowering the power density inside the fiber core is an alternative path to stretch the limit of SBS. Modern double-clad large-mode-area (LMA) micro-structured ytterbium (Yb) doped fiber technology, which supports single-mode operation with a mode field area of ~700 μm$^2$, can therefore be used to mitigate SBS [12]. However, it is very challenging to develop a fully monolithic LMA amplifier in the presence of air hole microstructures, and transitioning from standard-core (~10 μm) to large-core (>30 μm) diameters, which leads to severe mode-field mismatching. For this reason, the LMA fiber amplifiers are often used in a free-space configuration, diminishing the advantages of active optical fibers. However, thanks to the progress of fused-fiber technology, including the introduction of mode-field adapters (MFA) and LMA termination, these issues have been resolved [17]. Several groups have already demonstrated single-frequency high-power lasers up to several hundred watts based on Ytterbium-ion (Yb$^{3+}$) doped fiber amplifiers [11–16]. Despite these results, a high-

*Giorgio Santarelli

giorgio.santarelli@institutoptique.fr

[1]Laboratoire Photonique, Numérique et Nanosciences (LP2N), IOGS-CNRS-Université de Bordeaux, Talence 33405, France

[2]Azur Light Systems, Av. de la Canteranne, Pessac 33600, France

[3]ALPhANOV, 1 Rue François Mitterrand, Talence 33405, France

[4]Laboratoire ARTEMIS, UMR 7250 Université Côte d'Azur–CNRS–Observatoire de la Côte d'Azur, F-06304 Nice, France

[5]Amplitude-Systemes, 11 Avenue de la Canteranne 33600 Pessac, France

[6]Present address : MOE Key Laboratory of Fundamental Physical Quantities Measurement, School of Physics, Huazhong University of Science and Technology, 1037 Luoyu Rd., Wuhan 430074, China

power all-fiber laser operating at ultra-low amplitude noise reduction by using active stabilization has not yet been realized with reliable long-term operation and that is able to face the challenges of demanding, cutting-edge applications.

In this paper, we report on an all-fiber integrated 100 W fully-polarized monolithic Yb-doped fiber amplifier meeting the requirements of the GW detector Advanced Virgo (AdVIRGO). The monolithic architecture is suitable for low-noise operation and long-term reliability. This laser exhibits much less intensity noise compared to previously reported laser systems at comparable output power levels in free-running operation. Wideband laser intensity stabilization is based on active control of the pump laser diode driving current. This fast feedback loop (~1.3 MHz bandwidth) achieves a RIN level of -160 dBc/Hz (3 kHz –20 kHz) at its full output power.

## 2  All-fiber MOPA design

Our laser system is realized using a MOPA configuration as shown in Fig. 1. It consists of a seed source and two cascaded fiber amplifier stages. To achieve full fiber integration of our laser system, we use a fiber-pigtailed external cavity diode laser (ECDL, ALS IR1064 [18]) as the seed source. The ECDL emits 50 mW continuous wave at a center wavelength of 1064 nm. The measured linewidth of the ECDL is below 50 kHz [19]. A pre-amplifier stage is used to boost the seed power up to about 5 W. The gain medium consists in 3 m of double-clad polarization-maintaining Yb-doped fiber (Nufern 10/125) with a core diameter of 10 μm. Fiber-pigtailed isolators are placed between each stage to prevent backward beam propagation.

The final power amplifier is realized with about 2 m of polarizing large-mode-area Yb-doped fiber (PLMA-YDF, NKT Photonics DC-200/40-PZ-Yb). The microscope image of the fiber end face is shown in the inset picture of Fig. 1 (a). It features a core diameter of 40 μm, an inner clad diameter of 200 μm, and an outer clad diameter of 450 μm. The air-hole microstructure allows quasi single-transverse-mode operation with a mode field area of 760 μm$^2$. An air-clad structure enables pump power confinement with efficient core-pump overlap and high-power handling capabilities and nominal pump absorption of 10dB/m@976nm. Linearly-polarized beam propagation is ensured by the birefringence induced by the two boron-doped silica stress rods of the PLMA, typical Polarization Extinction Ratio (PER) is >20dB. The transition from the LMA to a standard ~10 μm double clad fiber is handled by a custom designed MFA, which exhibits insertion losses well below 1 dB for both pump and signal beams. The pre-amplifier output is spliced to a commercial high-power handling (300 W) (6+1)×1 polarization-maintaining fiber combiner. We inject into the combiner about 3 W of signal, four pumps of 60 W (nominal maximum output power) and an auxiliary pump of 10 W for intensity noise control purposes.

In order to eliminate back reflections and manage the high output power, the end face of the PLMA-YDF is spliced to an angle-polished endcap (8 degrees) as shown in Fig. 1 (b). The main amplifier module is assembled into a specially designed ring-shaped aluminum housing with an inner diameter of 24 cm, an outer diameter of 34 cm, and a thickness of 1.5 cm. For efficient heat removal at high output power, the aluminum module is attached to a small water-cooled optical bench, which is temperature regulated at 18 °C. After amplification, the output beam is collimated by a 30 mm focal lens with a Gaussian beam profile of 1.2 mm diameter. The unabsorbed pump is stripped by a dichroic mirror. The amplifier pump-to-signal optical conversion factor is about 67%, and the overall electrical consumption efficiency reaches nearly 12% (including pump diode supply and temperature stabilization, pre-amplifier consumption, electrical power supply conversion factors and fiber water cooling). The non-polarized clad modes are rejected by spatial filtering after transmission through a free-space polarization isolator.

## 3  100 W fiber amplifier output characterization

The amplified signal output power and isolator transmission power versus pump power is depicted in Fig. 2 (a). From the linear fit, the pump to signal conversion slope efficiency is 67%. A maximum signal output power of 118 W is achieved at a pump power of 173 W. The main obstacle involved with power scaling is fiber modal degradation produced by power transfer from the fundamental mode to higher order modes. This process is caused by a photo-darkening-induced thermo-optic index grating which leads to mode instability [20, 21]. The measured mode instability threshold is >120 W. Above such power, the laser output degrades into higher-order modes after long-term operation (a few hours). Here, for long-term laser reliability we operate the output power at 100 W. GW detection requires uninterrupted operation of the whole detector network for an extended period (several months, if not years). To test the system's reliability, we operated the laser system continuously while monitoring the output power after the isolator for over 25 days in a laboratory environment. This resulted in a mean power of 103.6 W and a standard deviation of 0.30 W, as shown in Fig. 2 (b). Optical spectra of the output signal at powers of 2 W, 20 W, and 100 W are shown in Fig. 2 (c). The spectrum was obtained at 2 W output without the main amplifier operation. The amplified spontaneous emission (ASE) suppression ratio is ~50 dB at 100 W output power (RBW: 0.05nm).

To prove that the system is GW detector compliant, a full characterization is required with respect to AdVIRGO specifications (see Ref. [22] for details). (i) The transverse modal contents in a laser beam can be analyzed by a scanning Fabry-Perot cavity while observing its transmission (or equally by scanning the laser frequency) [7, 12, 15]. However, the beam's fundamental Gaussian mode content can be directly measured by monitoring the reflected beam around the corresponding resonance. A sampled beam (~20 mW) was injected into a reference non-degenerate triangular cavity of finesse $F$=940. The reflected signal is depicted in Fig. 2 (d) and shows that the beam contains above 94% in the fundamental Gaussian mode, which is similar to other results that have been reported with this type of fiber [12]. (ii) The beam jitter is critical for a wide range of applications especially in GW detection. For instance, at the input of the GW detector a beam's jitter is converted into power fluctuations by a pre-mode cleaner cavity [22].
We have fully characterized beam tilt and shift noise power spectral densities, as displayed in Fig. 3. Here, tilt fluctuations were measured directly on a quadrant photodiode in the focal plane of a convex lens.

Shift fluctuations were inferred by measuring the jitter at different positions. Special care was taken in order to mitigate air turbulence effect. The measurement was done in a sealed environment on low power sampled beam. The main high-power beam was dumped far from the measurement area. Hence, the turbulence generated by the heat exchange of the beam dump with the surrounding air does not perturb the measurement process.

We clearly observe narrow-peaked structures in the 100 Hz– 1 kHz regions, which are attributed to mechanical resonances. The beam pointing results we obtained are within the AdVIRGO requirements (see Fig.3). The phase/frequency noise added by the amplifier stages has been proven to be negligible [23-25]. Nevertheless, we have performed a preliminary evaluation of the amplifier additive phase/frequency noise by a heterodyne Mach-Zehnder interferometer and the result is well below 1 Hz$^2$/Hz (10 Hz–10 kHz) consistent with previously published results.

## 4 Free-running intensity noise characterization and analysis

For a detailed characterization of fiber amplifier intensity noise, the free-running RIN is measured using a vector signal analyzer (HP89410A) at output powers of 2 W, 20 W, and 100 W (shown in Fig. 4). These measurements were obtained with an incident power of 20 mW on the photodiode. At Fourier frequencies lower than 100 Hz, the RIN is nearly independent of the output power and steeply increases toward lower frequencies. The increased noise is likely due to environmental perturbations such as air flow, mechanical vibrations, acoustic disturbances and chiller-induced vibrations. It has been shown that the low frequency (<100 kHz) intensity noise is due to pump laser RIN transfer to the laser by amplification gain dynamics [26–28]. The RIN of the pump diodes is nearly flat and its value decreases when the output power is increased. Therefore, a RIN of about -129 dBc/Hz can be achieved between 100 Hz and 100 kHz at 100 W laser output, which is in agreement with an analysis of the gain dynamics.

However, the RIN at high frequencies (>1 MHz) increases with the output power. The onset of SBS can be excluded because we use a short, large core fiber (with a beam diameter twice larger than for the previously reported 50 W fiber amplifier [19]) for which the SBS power threshold is estimated to be beyond 200 W which is well above the output power of our fiber amplifier. Moreover, using a Nd: YAG non-planar ring oscillator (NPRO) as a seed laser, a level of -160 dBc/Hz is recovered at 100 W for the frequency band 4 – 10 MHz. We are carrying out further investigations to explain the excess noise when the ECDL is used as a seed laser. Prior to this work, a comparison of free-running RIN (between a 160 W Nd: YAG amplifier and a 177 W injection locked laser) was reported in Refs. [8, 29]. Compared to such lasers our 100 W fiber amplifier shows substantially lower free-running RIN and meets AdVIRGO specifications (free-running requirements). It is worth noting that the integrated intensity noise from 10 Hz to 10 MHz, is about 0.024% (RMS) at 100 W output power.

## 5  Fiber amplifier intensity noise control

As shown above the high power fiber amplification adds significant intensity noise. In order to suppress this excess noise, two main techniques have been developed. One method utilizes a free space acoustic-optical modulator (AOM) to externally control the transmitted laser power. High handling power AOMs have significant time delay which severely limits the feedback bandwidth below 100 kHz [7, 30]. An alternative method is to directly control the current of the pump laser diodes (LD). In the early phases of laser intensity stabilization, such a scheme was applied to suppress the intensity noise of a multi-stage Nd: YAG amplifier [31–33]. The feedback bandwidth extending beyond 100 kHz remains challenging with high-power multi-mode pump laser diodes. In addition to bandwidth requirements, the intensity noise control of high power lasers requires a large dynamic range. For example, in Nd: YAG amplifiers the pump LDs are connected in series for efficient current supply [31–33]. A control current is added directly to the high current supply for all the pump LDs leading to a poor frequency response. Our system exhibits high pump-to-signal conversion efficiency and very low free-running intensity noise. Therefore, current requirements to suppress the intensity noise of our high-power fiber amplifier are strongly reduced.

To this aim, the current control is applied to the auxiliary multi-mode 60 W pump LD. The LD optical output power coefficient versus driving current is about 5 W/A. The use of a high speed (10 MHz full power bandwidth) buffer amplifier provides an output current in excess of 1 A (RMS). The frequency response of the servo control is shown in Fig. 5. The measurement is obtained by injecting a small chirped sine wave signal into the input of the servo actuator electronics and then directly modulating the auxiliary pump diode operating at an output power ~10 W. The output pump signal is detected by a fast photodiode and analyzed by a vector signal analyzer. Since the time delay of the photodiode is negligible, the measurement retrieves the frequency response of the servo system. The frequency response is flat up to 100 kHz and shows a 3 dB bandwidth of ~1 MHz. The time delay of the global transfer function amounts to less than 110 ns. In addition, there are no visible resonances or dips over the full span (10 Hz – 10 MHz). Compared to the AOM-based system used for intensity noise stabilization [7], our servo system exhibits much faster frequency response with a total time delay approximately ten times lower.

We use two identical InGaAs photodiodes (Perkin Elmer C30643), one for feedback control (in-loop) and one for independent noise measurement (out-of-loop). The 3 mm active diameter detectors have a measured response sensitivity of 0.76 A/W. Wideband low-noise transimpedance amplifiers follow the photodiodes with a measured input noise below 2 nV/(Hz)$^{1/2}$. The beam for detection is sampled by a wedge with 4% of total laser output power. For efficient photodiode illumination, the beam spot sizes are about 350 μm. The optical power incident on the detectors is adjusted by a polarizing beam splitter and then divided into the two paths by a 50: 50 non-polarizing beam splitter. The gain of each linear operational amplifier is carefully designed to avoid the saturation of feedback signals leading to a robust operation. With the servo loop closed, the measured RIN at 100 W output power are shown in the Fig. 6. The real stabilized intensity noise is characterized by the out-of-loop measurement. We measure a RIN of -160 dBc/Hz between 3 kHz and 20 kHz. Taking into account the noise of the transimpedance amplifier, the shot-noise limit can be expressed as,

$$S_q(dBc/Hz) = 10\log\left(2e/I_{out-loop} + (V_{in}/(I_{out-loop}R))^2 + 2e/I_{in-loop} + (V_{in}/(I_{in-loop}R))^2\right)$$

where $e$ is the elementary electric charge and $V_{in}$ is the input noise of transimpedance amplifier. The calculated shot-noise limit is -163 dBc/Hz, close to the measured value. The in-loop RIN starts to increase at frequencies higher than 10 kHz due to the servo gain profile. From 40 kHz, the in-loop RIN overlaps well with the out-of-loop RIN measurements. At frequencies higher than 700 kHz, the out-of-loop RIN crosses the free-running RIN measurement and follows with a servo bump at 1.3 MHz. Another small bump is found around 80 kHz. Several parasitic lines around 100 kHz are due to electronic-magnetic coupling (EMC) issues [26]. At frequencies below 1 kHz, the out-of-loop RIN steeply increases. For the out-of-loop measurement, we roughly suppress the free-running RIN by only 13 dB at frequencies below 100 Hz. The performance is substantially lower than the in-loop RIN measurement, which implies the existence of uncorrelated noise between the two detectors.

It is worth noting that our setup lacks a pre-mode cleaner cavity, spatial filtering and stray light minimization. Presumably, the beam jitter-to-RIN coupling through the non-uniform response of the photodiode surface is likely to be responsible of the excess intensity noise frequencies lower than 1 kHz [see Ref. 34 for further details]. This is consistent with previous comparisons of out-of-loop RIN measurements with and without a pre-mode cleaner cavity [32]. The effort to optimize the intensity stabilization below 1 kHz is beyond the scope on this work and will be pursued in the framework of the AdVIRGO project. Additionally, we evaluate the amplitude to phase conversion coefficient of the pump power fluctuations by slightly modulating the auxiliary pump diode (<0.01%) with a sine wave signal. We then acquire both the intensity by photodiode detection and the optical phase by processing the beat-note signal obtained with a heterodyne Mach-Zehnder to remove the seed laser frequency noise. The coefficient is of the order of 2 rad for 1% of power variation. Therefore, pump fluctuations and RIN feedback control have a negligible impact of optical phase noise.

# 6 Conclusion

In conclusion, we have developed a compact and robust 100 W ultra-low-noise all-fiber single-frequency laser. The fiber laser exhibits outstanding features (spatial mode content, long-term reliability, low beam jitter noise, low intensity noise) thanks to which the system is compliant with AdVIRGO requirements provided a low phase noise seed laser. By means of a direct current control circuit actuating on an auxiliary pump LD, a maximum RIN suppression of 30 dB is demonstrated with broadband feedback bandwidth, resulting in 1.3 MHz servo bandwidth and a close to shot-noise-limited RIN performance of -160 dBc/Hz (3–20 kHz). To the best of our knowledge, the feedback bandwidth beyond 1 MHz was achieved for the first time at an output power level of 100 W.

Our future work targets a further increasing output power available for GW detectors using the coherent combing scheme [35] and the investigation of novel LMA fibers. Using the power stabilization techniques developed for GW detection experiments [7, 10], we anticipate that it will be possible to meet the high power and low noise requirements for present day [36, 37] and future GW detectors.

**Acknowledgements**

This work is supported Agence Nationale de la Recherche (ANR) (ANR14 LAB05 0002 01) and Conseil Régional d'Aquitaine (2014-IR60309-00003281); the author J. Zhao acknowledges Post-doctoral scholarship grant from La Fondation Franco-Chinoise pour la Science et ses Applications (FFCSA); European Gravitational Observatory (EGO). We thank Dr. Benoit Gouhier and Dr. Sergio Rota-Rodrigo for their comments on the paper.

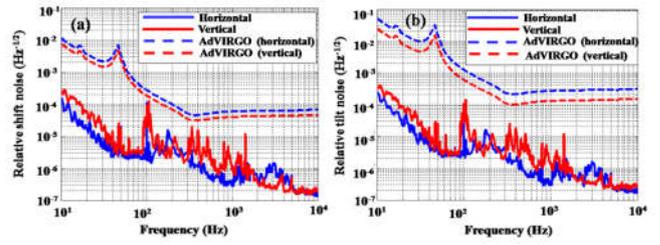

**Fig. 3.** Measured relative beam pointing spectral densities of a) shift and b) tilt, and Advanced VIRGO (AdVIRGO) requirements. (AdVIRGO requirements were calculated assuming a four mirrors pre-mode cleaner ring cavity (1.6 m of round trip) with a Finesse of 120 before injecting the beam into the mode cleaner cavity).

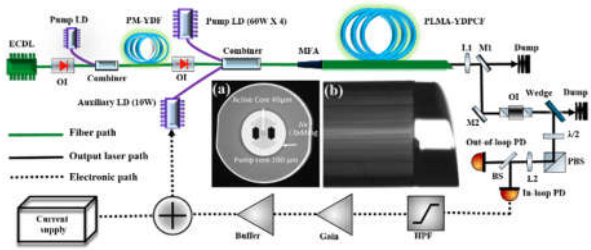

**Fig. 1.** Schematic of the ultra-low intensity noise all-fiber laser. ECDL: external-cavity diode laser; OI: optical isolator; MFA: mode field adapter; L: lens; M: mirror; PD: photodiode; PM-YDF: polarization-maintaining Yb-doped fiber; PLMA-YDPCF: polarizing large-mode area Yb-doped photonic crystal fiber. Inset (a) shows the microscope image of the fiber end face. Inset (b) is an image of the PLMA end cap.

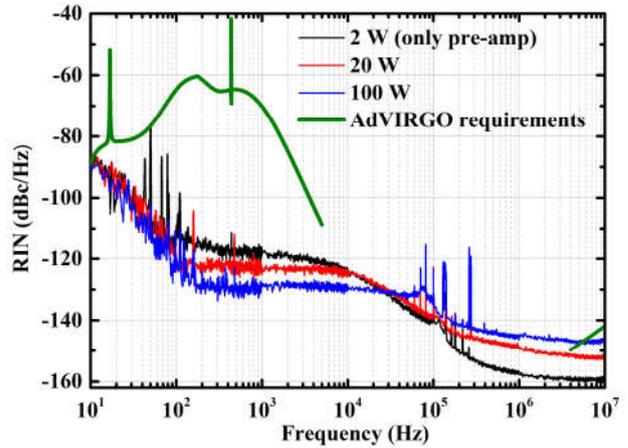

**Fig. 4** Free-running RIN characterization at different power level of 2 W (black curve), 20 W (red curve), 100 W (blue curve) and AdVIRGO requirements for free running laser (green curve).

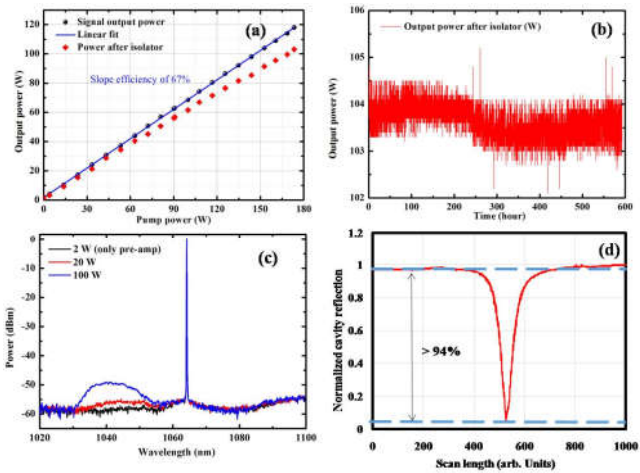

**Fig. 2.** (a) Output power versus pump power before and after optical isolator. (b) Long-term fiber amplifier output power stability after the isolator. (c) Output spectrum at a power level of 2 W, 20 W, and 100W (RBW: 0.05nm). (d) Cavity reflection signal.

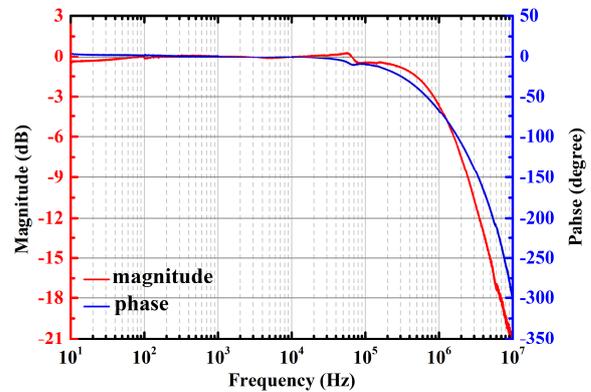

**Fig. 5.** Measured frequency response of the servo system (magnitude & phase).

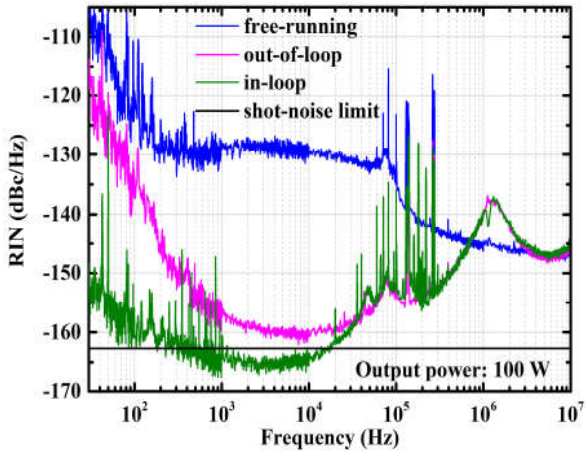

**Fig. 6.** Performance of the servo loop measured by in-loop (green trace) and out-of-loop (pink trace) photodiodes at 100 W output power. Also shown the free-running RIN (blue trace) and the calculated photocurrent shot-noise limit (black line).